\documentclass[aps,twocolumn]{revtex4}
\usepackage{bm}
\usepackage{epsfig}
\usepackage{amsmath}
\usepackage{amssymb}
\usepackage{graphicx}

\newcommand{\sign}{\mathop{\rm sign}\nolimits}

\begin{document}

\title{Suppression of spin beats in magneto-oscillation
phenomena\\ in two-dimensional electron gas}

\author{N.S.~Averkiev}
\author{M.M.~Glazov}
\author{S.A.~Tarasenko}
\affiliation{A.F.~Ioffe Physico-Technical Institute of the RAS,
194021 St.~Petersburg, Russia}


\begin{abstract}
Theory of magneto-oscillation phenomena has been developed for
two-dimensional electron systems with linear-in-$\bm{k}$ spin
splitting. Both Dresselhaus and Rashba contributions are taken
into account. It has been shown that the pattern of the
magneto-oscillations depends drastically on the ratio between the
above terms. The presence of only one type of the $\bm{k}$-linear
terms gives rise to the beats, i.e. two close harmonics
corresponding to the spin-split subbands. However, if the
strengths of both contributions are comparable, the third
(central) harmonics appears in the spectrum of the
magneto-oscillations. For equal strengths of the contributions,
only the central harmonic survives, and the oscillations occur at
a single frequency, although the $\bm{k}$-linear terms remain in
the Hamiltonian. Such suppression of the spin beats is studied in
detail by the example of the Shubnikov-de Haas effect.
\end{abstract}

\maketitle

\section{Introduction}
Spin-dependent transport phenomena in two-dimen\-sional electron
systems is of broad interest at present time. The peculiar
property of the two-dimensional systems based on quantum wells is
linear in the wave vector $\bm{k}$ spin-dependent terms in the
effective Hamiltonian~\cite{Bychkov,D'yakonov}. These terms are
caused by spin-orbit interaction that couples spin states and
space motion of conduction electrons and governs the wide class of
spin phenomena. Among them are spin relaxation, spin transport
controlled with an external electric field by the Rashba effect,
circular photogalvanic and spin-galvanic effects, electric
current-induced spin orientation and precession, intrinsic spin
Hall effect, etc.

Experimentally one of the most efficient methods for determination
of the spin splitting in 2D conducting structures is measurements
of oscillations of the magnetoresistivity (Shubnikov-de Haas
effect). The quantum oscillations are highly sensitive to the fine
structure of the energy spectrum of carriers, so that even small
spin splitting, small as compared to the Fermi energy but
commensurable with the energy distance between Landau levels,
qualitatively modifies the oscillation behavior. The linear in the
wave vector terms in the effective Hamiltonian remove the
degeneracy in the carrier spectrum. In a magnetic field, the spin
splitting at the Fermi surface gives rise to the oscillations with
close frequencies, i.e. to beats~\cite{Bychkov}. Such a behavior
was observed and attributed to the zero-field spin splitting in 2D
electron gas under study of the Shubnikov-de Haas oscillations in
InAs/GaSb~\cite{Luo,Ivanov}, InGaAs/InAlAs~\cite{Das,Nitta,Cui},
InGaAs/InP~\cite{Engels}, AlGaAs/GaAs~\cite{Ramvall},
InAs/AlSb~\cite{Heida}, and InGaN/GaN-based~\cite{Lo} structures,
and microwave radiation-induced magnetoresistance oscillations in
GaAs/AlGaAs heterostructures~\cite{Mani}. The analysis of the
oscillations was applied for determination of the spin splitting
at the Fermi level. However, recently it was pointed out that the
simple analysis of the beating pattern may lead to an incorrect
conclusion on the spin splitting~\cite{Keppeler,JETPL}. In
particularly, it was shown that the $\bm{k}$-linear contribution
induced by heteropotential asymmetry (Rashba term) and the
contribution originated from the cubic terms of a bulk material
lacking inversion center (Dresselhaus term) interfere in the
magneto-oscillation phenomena~\cite{JETPL}. The presence of only
one type of the linear terms gives rise to the beats. However, if
the strengths of both contributions are equal, the oscillations
occur only at a single frequency and the beats disappear, although
the $\bm{k}$-linear terms remain in the Hamiltonian.

In this communication we present a theory of magneto-oscillation
phenomena in two-dimensional electron gas in the presence of
$\bm{k}$-linear spin splitting. Both $\bm{k}$-linear contributions
originated form a heteropotential asymmetry and the lack of the
inversion center in a bulk semiconductor are taken into account.
We take the Shubnikov-de Haas effect as an example to study the
suppression of the spin beats in detail. The magnetoresistance is
calculated for arbitrary ratio between the Rashba and Dresselhaus
contributions. The Zeeman splitting of electronic levels is
neglected for simplicity, since it is small compared to the
spacing between the Landau levels in the magnetic field
perpendicular to the quantum well plane in the majority of
semiconductor structures based on the III-V compounds.

\section{Theory}
Generally, appearance of the $\bm{k}$-linear terms is connected
with reduction of the system symmetry as compared to the bulk
material. In (001)-grown quantum wells based on
zinc-blende-lattice semiconductors, there are two types of the
linear contributions to the effective Hamiltonian of 2D electrons.
First, they originate from the lack of the inversion center in the
bulk material (BIA term)~\cite{D'yakonov} (or from the asymmetry
of the chemical bonds at interfaces (IIA term)~\cite{IIA}). This
so-called Dresselhaus contribution has the form
\begin{equation}\label{dress}
\mathcal H_{D} = \alpha (\sigma_x k_y + \sigma_y k_x) \:,
\end{equation}
where $\sigma_i$ ($i=x,y$) are the Pauli matrices and the axes $x$
and $y$ are assumed to be parallel to the crystallographic axes
$[1 \bar1 0]$ and $[110]$, respectively. Second, a linear
contribution can be induced by the heterostructure asymmetry
unrelated to the crystal lattice (SIA or Rashba
term~\cite{Bychkov})
\begin{equation}\label{rashba}
\mathcal H_{R} = \beta (\sigma_x k_y - \sigma_y k_x) \:.
\end{equation}
Experimental data evidence that the intensities of the Dresselhaus
and Rashba terms can be comparable in real 2D structures based on
quantum wells~\cite{Knap,Jusserand,SGE_opt}. Furthermore, the
relative intensities of the contributions can be controlled tuning
the heteropotential asymmetry with an external electric field
normal to the quantum well plane.

In two-dimensional systems small magneto-oscillations occur in
moderate magnetic fields, when $\omega_c \tau \leq 1$, where
$\omega_c=eB/m^*c$ is the cyclotron frequency, $\tau$ is the
scattering time, $e$ is the elementary charge, $\bm{B}$ is the
applied magnetic field, $m^*$ is the effective electron mass, and
$c$ is the light velocity. The corresponding parameter that
determines the amplitude of the oscillations appears to be
$\exp{(-\pi/\omega_c\tau)}$~\cite{Ando,Isihara,AGTW,FTT}. We
assume that the inequality $E_F \tau /\hbar \gg 1$ providing good
conductivity is fulfilled, and the spin splitting at the Fermi
surface is much less than the Fermi energy $E_F$ but exceeds the
spacing between the Landau levels, $\hbar\omega_c \ll
\sqrt{\alpha^2+\beta^2}\:k_F \ll E_F $, where $k_F$ is the Fermi
wave vector. Then, in the self-consistent Born approximation, the
single-particle electron Green's function under electron
scattering by short-range defects has the form
\begin{equation}\label{green}
\hat{ \mathcal G}_\varepsilon (\bm r, \bm r') = \sum_{n,k_y}
\frac{\Psi_{n k_y}(\bm r) \Psi_{n k_y}^\dag(\bm r')}{\varepsilon
+E_F - E_n - X_\varepsilon} \:,
\end{equation}
where $\Psi_{n k_y}(\bm r)$ are the spinor wave functions for an
electron subjected to the magnetic field ${\bf B}\parallel z$ with
the vector potential taken in the Landau gauge, $\bm A =
(0,Bx,0)$, $E_n$ are the electron levels, $X_\varepsilon$ is the
self-energy part of the Green's function, $k_y$ is the
$y$-component of the wave vector, and $n$ is the quantum number
enumerating both the Landau levels and the spin states. We note,
that the Green's function~(\ref{green}) is a $2\times 2$ matrix in
the spin indices. The self-energy part obeys, for the large
numbers $n\sim 2 E_F/\hbar\omega_c$, the following equation
\begin{equation}\label{selfenergy}
X_\varepsilon = \frac{\hbar^2\omega_c}{4\pi\tau} \sum_n
\frac{1}{\varepsilon +E_F - E_n - X_\varepsilon} \:.
\end{equation}
Quantum oscillations are determined, to the first order in the
parameter $\exp{(-\pi/\omega_c\tau)}$, by the imaginary part of
the self-energy. The real part of $X_{\varepsilon}$ can be assumed
to be included in the Fermi energy $E_F$. Within this accuracy the
solution of the Eq.~(\ref{selfenergy}) has the form
\begin{equation}\label{selfenergy1} X_{\varepsilon} =
-{\rm i}\frac{\hbar^3\omega_c}{8\pi\tau} \sum_n
\frac{\sign{\varepsilon}}{(\varepsilon +E_F - E_n)^2 +
(\hbar/2\tau)^2} \:.
\end{equation}

The eigen wave functions $\Psi_{n k_y}(\bm r)$ and the energies
$E_n$ are determined from the solution of the Schr\"{o}dinger
equation, with the terms $\mathcal H_{\rm D}$ and $\mathcal H_{\rm
R}$ being included in the Hamiltonian. In the presence of only one
type of the linear terms (either Dresselhaus or Rashba), electron
states in the magnetic field can be conveniently characterized by
two indices $n=(m,s)$, where the index $s=\pm$ denotes spin
states. Then the electron energies are given by~\cite{Bychkov}
\begin{equation}\label{only1}
E_{m, \pm }  = \hbar \omega _c  \, m \pm \sqrt { 4 \alpha^2 \,m/
\lambda^2_B + (\hbar\omega_c)^2/4 } \:,
\end{equation}
where $\lambda_B=\sqrt{\hbar c/eB}$ is the magnetic length. It is
the splitting $E_{m,+} - E_{m,-} = 4\alpha \sqrt{m}/\lambda_B
\approx 2\alpha k_F$ that gives rise to the beats in
magneto-oscillations~\cite{Bychkov,JETPL,Wang,Langenbuch}. In the
other limiting case, $|\alpha| = |\beta|$, the orbital motion and
the spin states can be separated, and the electron levels become
double-fold degenerate~\cite{JETPL}
\begin{equation}\label{equal}
E_{m,\pm} = \hbar \omega_c(m+1/2) - 2m^*\alpha^2/\hbar^2.
\end{equation}
The $\bm{k}$-linear terms, although being present in the
Hamiltonian, do not lead to the splitting of the Landau levels,
and the beats in the magneto-oscillations do not occur.

\begin{figure}[b]
\includegraphics[width=0.47\textwidth]{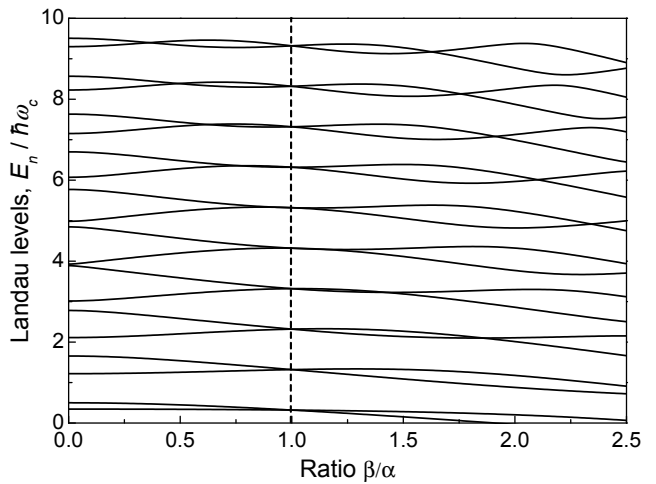}\\
\caption{The positions of the Landau levels at different ratio
$\beta/\alpha$ for the fixed magnetic field and $m^* \alpha^2/(2
\omega_c \hbar^3)=0.045$.}\label{fig1}
\end{figure}
For the arbitrary strengths of the Dresselhaus and Rashba terms
the electron spectrum in the magnetic field can be calculated
numerically. Fig.~\ref{fig1} presents the dependence of the
arrangement of the Landau levels on the ratio $\beta / \alpha$
calculated for the value $m^* \alpha^2/(2 \omega_c \hbar^3)=0.045$
that is an estimation for GaAs/AlGaAs structures in the magnetic
field 1~T. One can see that the dependence of the energy spectrum
on $\beta/\alpha$ is rather complicated. At arbitrary ratio
$\beta/\alpha$ the electron states are non-degenerate, and one can
expect the complicated pattern of the Shubnikov-de Haas
oscillations. However, when $\alpha=\beta$ the spectrum is a set
of the equidistant Landau levels with the energy distance $\hbar
\omega_c$, and the oscillations occur at a single frequency.

The Green's function allows us to calculate various kinetic and
thermodynamic coefficients. We demonstrate it taking the
Shubnikov-de Haas effect as an example.

The conductivity tensor at a frequency of the external electric
field $\omega$ is expressed in the terms of the Green's function
by
\begin{equation}\label{sigmaab}
\sigma_{ij}(\omega) = {\rm i} \frac{Ne^2}{m^*\omega}\,\delta_{ij}
+ \frac{1}{\omega}\, {\rm Tr} \int \int d\bm{r} d\bm{r}'
\end{equation}
\vspace{-0.5cm}
\[
\hspace{1.5cm} \times \int\limits_{-\infty}^{\infty}
\frac{d\varepsilon}{2\pi} [\hat{J}_{i}(\bm{r})\, \hat{ \mathcal
G}_{\varepsilon+\hbar\omega} (\bm r, \bm r')]\,
[\hat{J}_{j}(\bm{r}')\, \hat{ \mathcal G}_{\varepsilon} (\bm r',
\bm r)] \:.
\]
Here $N$ is the carrier concentration, $\delta_{ij}$ is the
Kroneker symbol, $\hat{J}_{i}$ is the current density operator
that, in general case, is a $2\times 2$ matrix in the spin
indices, $i,j=x,y$ are the in-plane co-ordinates, and the trace
${\rm Tr}$ implies summation over the spin indices. Below we are
interested in the static conductivity, and therefore consider the
frequency $\omega$ to be a small value and reduce it to zero in
the final expressions.

The analysis shows that the oscillations of the longitudinal
conductivity are determined by the integral $\int_{-\hbar\omega}^0
d\varepsilon$ in Eq.~(\ref{sigmaab}), where the self-energy parts
$X_{\varepsilon}$ and $X_{\varepsilon+\hbar\omega}$ have the
opposite signs~\cite{Ando,Isihara,AGTW,FTT}. The other regions of
the integration over $\varepsilon$ in Eq.~(\ref{sigmaab}),
$-\infty<\varepsilon<-\hbar\omega$ and $0<\varepsilon<\infty$,
compensate the gauge term ${\rm i} N e^2/m^*\omega$. To the first
order in the parameter $\exp{(-\pi/\omega_c\tau)}$ the
longitudinal conductivity is given by
\begin{equation}\label{sigmaxx}
\sigma_{xx} = \frac{m^*\omega_c}{4\pi^2} \sum_{nn'}
\frac{|(J_x)_{nn'}|^2 }{(E_F - E_n - X_{0+})(E_F - E_{n'} -
X_{0-})} \:.
\end{equation}
where $(J_x)_{nn'}$ are the matrix elements of the current density
operator, and $X_{0\pm}$ are the limits of the self-energy parts
at $\varepsilon\to 0$, $X_{0\pm} = \lim\limits_{\varepsilon\to
0\pm} X_{\varepsilon}$. The oscillations of the Hall conductivity
$\sigma_{xy}$ are determined by the whole range of the integration
over $\varepsilon$ in Eq.~(\ref{sigmaab}), from $-\infty$ to
$+\infty$. To the first order in the parameter
$\exp{(-\pi/\omega_c\tau)}$ the Hall conductivity is derived to be
\begin{equation}\label{sigmaxy}
\sigma_{xy} = \frac{m^*\omega_c}{4\pi^2} \sum_{nn'} (J_x)_{n'n} \,
(J_y)_{nn'}
\end{equation}
\vspace{-0.6cm}
\[
\hspace{-2.3cm} \times \left[ \frac{1}{(E_F - E_n - X_{0+})(E_F -
E_{n'} - X_{0-})} \right.
\]
\vspace{-0.4cm}
\[
+\, 2{\rm i} \frac{ \arctan{[2\tau(E_F - E_n)/\hbar}] -
\arctan{[2\tau(E_F - E_{n'})/\hbar]} }{(E_n - E_{n'})^2}
\]
\vspace{-0.5cm}
\[
\hspace{0.3cm} \left. -\, \frac{E_{n'} - E_n + {\rm
i}\hbar/\tau}{(E_{n'} - E_n)(E_F - E_n + \rm i\hbar / 2\tau)(E_F -
E_{n'} - \rm i \hbar/ 2\tau)} \right] \:.
\]
Eqs.~(\ref{sigmaxx})~and~(\ref{sigmaxy}) describe the
magnetoconductivity of the two-dimensional electron system in the
regime of the Shubnikov-de Hass oscillations. The main
contribution to the sums~(\ref{sigmaxx},\ref{sigmaxy}) is
introduced by the terms with the energies $E_n,E_{n'}$ close to
the Fermi surface. Therefore on analytical calculation, when the
summation over $n$ is replaced by an integration with the Poisson
formula, the integration is performed over residua of the Green's
function. It allows one to avoid divergency at $n \rightarrow
\infty$ in $\sigma_{xx}$ and $\sigma_{xy}$ that is caused by the
increase of the matrix element $(J_{i})_{nn'}$ with
$n$~\cite{AGTW,FTT}. The similar summation ``over residua'' in
numerical calculations can be performed substituting the matrix
elements $(J_{i})_{nn'}$ by $\sqrt{2E_F/(n\hbar\omega_c)}
\,(J_{i})_{nn'}$.

\section{Results and discussion}

\begin{figure}[t]
\includegraphics[width=0.48\textwidth]{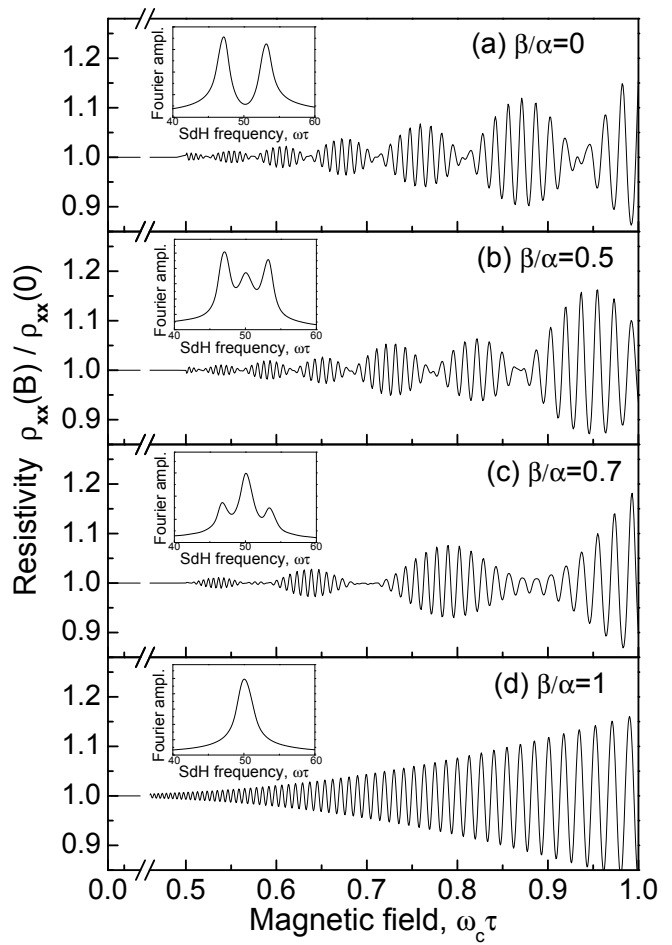}\\
\caption{The magnetic-field dependence of the resistivity
$\rho_{xx}$ for different ratios of Rashba to Dresselhaus terms,
$\beta/\alpha= 0, 0.5, 0.7, 1$, at fixed Fermi energy $E_F
\tau/\hbar = 50$ and fixed Dresselhaus splitting $\alpha
k_F\tau/\hbar  =  3$. The insets present the Fourier spectra of
the corresponding dependencies $\rho_{xx}$ on
$1/\omega_c\tau$.}\label{fig_osc}
\end{figure}
Figure~\ref{fig_osc} displays the magnetic-field dependence of the
resistivity
\[
\rho_{xx} = \frac{\sigma_{xx}}{\sigma_{xx}^2+\sigma_{xy}^2}
\]
calculated numerically following
Eqs.~(\ref{sigmaxx},\ref{sigmaxy}) for various ratios of Rashba to
Dresselhaus spin-splitting constants, $\beta/\alpha$. The ratio
$\beta/\alpha$ was varied modifying the strength of Rashba term,
while the Dresselhaus term was kept constant. The insets in the
Figures present the Fourier spectra of the corresponding
dependencies $\rho_{xx}$ on $1/\omega_c\tau$.

For the case of only one type of the $\bm{k}$-linear terms ($\beta
=0$, Fig.~\ref{fig_osc}a) the Shubnikov-de Haas oscillations
demonstrate the pronounced beats. The phase of the oscillations
reverses at the node points. As shown in the inset, the spectrum
of the oscillations consists of two harmonics, with the spectral
positions corresponding to the energies of the spin subbands, $E_F
\pm \alpha k_F$. The oscillating part of the resistivity is
described analytically by $\exp{(-\pi/\omega_c\tau)} \cos{(2\pi
\alpha k_F/\hbar\omega_c)} \cos{(2 \pi E_F /\hbar\omega_c)}$.

In the presence of both Rashba and Dresselhaus contributions of
comparable strengths the pattern of the magneto-oscillations
modifies (Figs.~\ref{fig_osc}b,~\ref{fig_osc}c). As can be seen in
the insets the central peak arises in the Fourier spectrum. For
the parameter set presented in the Figure caption, the
oscillations at $\beta/\alpha=0.5$ (Fig.~\ref{fig_osc}b)
demonstrate the beat pattern similar by sight to that at $\beta=0$
(Fig.~\ref{fig_osc}a), although the Fourier spectrum shows clearly
the mixture of three harmonics. Further increase of the ratio
$\beta/\alpha$ to 1 results in the noticeable modification of the
magneto-oscillations (Fig.~\ref{fig_osc}c). The beats become
irregular, the reverse of the oscillation phase at nodes
disappears, and with further approaching of $\beta/\alpha$ to 1
the nodes disappear.

In the case of the equal strengths of the Rashba and Dres\-selhaus
terms, the beats vanish completely and the
mag\-ne\-to-oscil\-la\-tions oc\-cur at a sin\-gle fre\-quen\-cy,
al\-though the $\bm{k}$-linear spin splitting remains for the most
of the directions of the wave vector $\bm{k}$. In this particular
case the oscillations are described by
$\exp{(-\pi/\omega_c\tau)}\cos{(2\pi E'_F/\hbar\omega)}$, where
$E'_F=E_F+2m^* \alpha^2/\hbar^2$ is the Fermi level measured from
the bottom of the spin subbands~\cite{JETPL}.

\begin{figure}[t]
\vspace{0.2cm}
\includegraphics[width=0.47\textwidth]{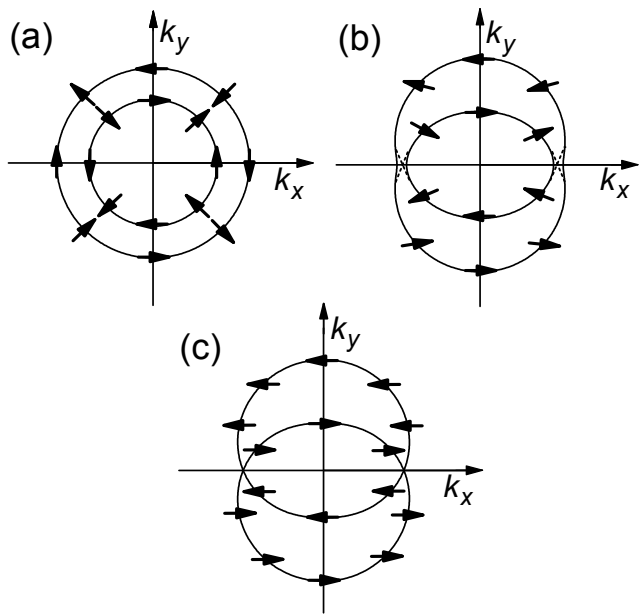}\\
\caption{Fermi contours for electrons in (001)-grown quantum wells
in zero magnetic field in the presence of (a) Dresselhaus term
only, $\beta=0$, (b) Rashba and Dresselhaus terms, (c) both
contributions of the equal strengths, $\beta/\alpha=1$. Arrows
indicate the orientation of spins. Dotted lines in Fig.~(b) show
possible transitions between the closely located
orbits.}\label{orbits}
\end{figure}

Such a behavior of the magneto-oscillations at various
$\beta/\alpha$ can be understood qualitatively considering the
simple semiclassical picture. In the framework of the
semiclassical treatment the frequencies of the
magneto-oscillations correspond to the Borh-Sommerfeld
quantization of an electron motion over the classical cyclotron
orbits (Fig.~\ref{orbits}). For the case when the spin-orbit
splitting exceeds the cyclotron energy $\hbar\omega_c$, the
electrons move over the cyclotron orbits, with the spins
adiabatically oriented parallel or antiparallel to the effective
magnetic field $\bm{B}_{so}$ induced by the spin-orbit
coupling~\cite{Keppeler}. The orbits corresponding to the spins
oriented along and opposite to $\bm{B}_{so}$ are split in
$\bm{k}$-space. In the presence of the Dresselhaus term only
(Fig.~\ref{orbits}a), the electron spectrum in zero external field
has the form
\[
E_{\bm{k},\pm} = \hbar^2 k^2 /2m^* \pm \alpha k \:.
\]
Quantization of cyclotron orbits corresponding to the spins
oriented parallel and antiparallel to $\bm{B}_{so}$ gives rise to
two harmonics in the magneto-oscillations, i.e. the beating
pattern.

In the presence of both Rashba and Dresselhaus contributions of
comparable strengths~(Fig.~\ref{orbits}b) the electron spectrum
becomes strongly anisotropic~\cite{Andrada}. The condition of
adiabatic spin orientation along the cyclotron orbit is broken in
certain points in the $\bm{k}$-space where the spin splitting is
small. In the vicinity of the points the electron spin does not
follow adiabatically the effective field $\bm{B}_{so}$, and
electron transitions from one spin orbit to the other become
possible. The magnetic breakdown occurs, i.e. the electron
tunneling between the close cyclotron orbits~\cite{breakdown}. The
transitions corresponding to the magnetic breakdown between the
orbits are shown in Fig.~\ref{orbits}b with dotted lines.
Quantization of the cyclotron motion with the transitions between
the orbits results in appearance of the third (central) harmonics
in the magneto-oscillations. The increase of the ratio
$\beta/\alpha$ to 1 leads to increase of the probability of the
tunneling between the orbits, i.e. to enhancement of the central
harmonic and depression of the low- and high-frequency harmonics.

In the case of the equal strengths of the Rashba and Dresselhaus
terms, the Fermi surface consists of two identical circles shifted
relative to each other in the $\bm{k}$-space. For $\alpha=\beta$
the circles are shifted along $k_y$ and characterized by the spin
states $| \pm 1/2 \rangle$ onto the $x$-axis~(Fig.~\ref{orbits}b).
In this particular case electrons within each spin subband are
quantized identically in the external magnetic field, and only one
(central) harmonics remains in the spectrum of the
magneto-oscillations. The magnetic breakdown between the subbands
does not occur, because the electron spins in the subbands are
opposite directed. The electron spin initially oriented parallel
or antiparallel to $\bm B_{so}$ keeps the orientation during the
electron motion over the cyclotron orbit because the effective
magnetic field is collinear to $x$-axis independent of the wave
vector.

The situation when the Rashba contribution to the spin splitting
is larger than the Dresselhaus term, $\beta/\alpha>1$, is similar
to the case of $\beta/\alpha<1$ considered above. Increase of the
ratio $\beta/\alpha$ from 1 leads to appearance of the three
harmonics in the spectrum of the magneto-oscillations again. When
the Rashba contribution dominates, only two harmonics
corresponding to the spin-split subbands remain in the spectrum
and the pattern of the magneto-oscillations is similar to that
presented in Fig.~\ref{fig_osc}a.

This work was supported by the RFBR, the INTAS, programs of the
RAS and the Russian Ministry of Science and Education, and
Foundation ``Dynasty'' - ICFPM.

\end{document}